# Complex worlds from simple rules?

John Casti's review "Science is a computer program" (*Nature* **417**, 381-382; 2002; see also News Feature "What kind of science is this?" *Nature* **417**, 216-218; 2002) of Stephen Wolfram's recent book[1] raises an intriguing question: Can the apparent complexity we observe in the real world be generated from simple initial conditions via simple, deterministic rules, much like the rules governing the evolution of Wolfram's cellular automata?

Remarkably, quantum theory supplies a partial answer to this question: unless we are willing to allow the possibility of faster-than-light communication, the observed Universe *could not* have evolved from simple initial conditions with simple, deterministic rules. This result is a corollary of two fundamental observations about our Universe: the existence of entangled quantum states for spatially separated composite systems, and the fact that the Universe is large and expanding.

In a quantum-mechanical world, one way to generate complexity is to perform measurements that project a fixed quantum state onto a non-orthogonal subspace in Hilbert space. For example, given an ensemble of electron spins all prepared in the "up" state along the *z*-axis, measuring them along the *x*-axis yields a probabilistic sequence of bits (0's and 1's depending on whether the measurement result is "left" or "right") whose complexity can be quantified with the notion of algorithmic (Kolmogorov) complexity[2] (a precise notion of randomness). Now, a world that evolves out of simple initial conditions via simple, deterministic rules must necessarily have small algorithmic complexity. However, the Einstein-Podolsky-Rosen-Bell correlations inherent in any entangled state guarantee that there exists a lower bound on the algorithmic complexity of quantum measurement sequences: violating this lower bound would allow spacelike-separated observers who share entanglement to communicate with each other[3] (spacelike communication in turn violates relativistic causality since, when combined with Lorentz invariance, it allows signalling into the past). This complexity lower bound is proportional to the spatial separation between the entangled sub-systems.[3] Consequently, the very large present size of the Universe guarantees a very large lower bound, and the expansion of the Universe guarantees that the bound will grow indefinitely, becoming even larger as the Universe becomes older.

If one must maintain faith in "simple rules" (as most physicists would), one needs to either abandon "simple initial conditions," or else introduce an element of non-deterministic evolution via an "oracle" that injects randomness at each quantum measurement step. Otherwise, simple, deterministic evolution rules starting from simple initial conditions cannot possibly produce the large algorithmic complexity inherent in our quantum world.

**Ulvi Yurtsever**

*Quantum Computing Technologies Group, Jet Propulsion Laboratory, California Institute of Technology, 4800 Oak Grove Drive, Pasadena, California 91109, USA.*